\documentstyle[aps,epsf]{revtex}
\def\Journal#1#2#3#4{{#1}{\bf #2}, #3 (#4)}

\def\NC{\em Nuovo Cimento }
\def\AP{\em Ap. J. }
\def\APS{\em Ap. J. Suppl. }
\def\NPB{{\em Nucl. Phys.} B}
\def\NPPS{\em Nucl. Phys. Proc. Suppl. }
\def\N{\em Nature }
\def\PLB{{\em Phys. Lett.} B}
\def\PRL{\em Phys. Rev. Lett. }
\def\PRD{{\em Phys. Rev.} D}
\def\PR{\em Phys. Rep. }
\def\RMP{\em Rev. Mod. Phys. }
\def\SJNP{\em Sov. J. Nucl. Phys. }

\def\ZPC{{\em Z. Phys.} C}
\def\ibid{\it ibid. }

\def\be{\begin{equation}}
\def\ee{\end{equation}}
\def\bea{\begin{eqnarray}}
\def\eea{\end{eqnarray}}

\begin{document}

\title
{ Cosmological Nucleosynthesis 
and Active-Sterile Neutrino Oscillations\\
with Small Mass Differences: The Nonresonant Case
}
\author
{ D. P. Kirilova}
\address
{
Teoretisk Astrofysik Center\\
Juliane Maries Vej 30, DK-2100, Copenhagen\\
and \\
Copenhagen University,
Niels Bohr Institute,
Blegdamsvej 17\\  DK-2100, Copenhagen, Denmark
\footnote{Permanent address:
 Institute of Astronomy, Bulgarian Academy of Sciences,\\
blvd. Tsarigradsko Shosse 72, Sofia, Bulgaria\\
E-mail:$dani@libra.astro.acad.bg$}
}

\author
{M. V. Chizhov}
\address
{
Centre for Space Research and Technologies, Faculty of Physics,\\
University of Sofia, 1164 Sofia, Bulgaria\\
E-mail: $mih@phys.uni$-$sofia.bg$}

\maketitle

\begin{abstract}
We study the nonresonant oscillations between left-handed
electron neutrinos $\nu_e$ and nonthermalized sterile neutrinos
$\nu_s$ in the early Universe plasma. The case
 when $\nu_s$ do not thermalize till 2 MeV and the oscillations become 
effective after $\nu_e$ decoupling is discussed. As far as for this model 
the rates of expansion of the Universe, neutrino oscillations and neutrino
 interactions with the medium may be comparable, we  have analysed the kinetic 
equations for neutrino density matrix, accounting {\it simultaneously} 
for all processes. The evolution of neutrino ensembles was described 
numerically by integrating the kinetic equations 
for the neutrino density matrix in {\it momentum} space for small 
mass differences $\delta m^2 \le 10^{-7}$ eV$^2$.  
 This approach allowed us 
to study precisely the evolution of the neutrino number densities,
energy spectrum distortion and the asymmetry between neutrinos and
antineutrinos due to oscillations for each momentum mode.
 
We have provided a detail numerical study of the influence of the
 nonequilibrium $\nu_e \leftrightarrow \nu_s$ oscillations on the 
primordial production of  $^4\! He$. The exact kinetic approach
enabled us to calculate the effects of neutrino population depletion,
    the  distortion of the neutrino 
spectrum and the generation of neutrino-antineutrino asymmetry 
on the kinetics of neutron-to-proton 
transitions during the primordial nucleosynthesis epoch and 
correspondingly on the cosmological  $^4\! He$ production. 

 It was shown that the neutrino population depletion and spectrum distortion  
play an important role. The asymmetry effect, in case the lepton asymmetry 
is accepted initially equal to the baryon one, is proved to be 
negligible for the discussed range of $\delta m^2$.
Constant helium contours in 
$\delta m^2$ - $\vartheta$ plane were calculated. Thanks to the
exact kinetic approach 
more precise cosmological constraints on the 
mixing parameters were obtained.
 
\end{abstract}

\ \\

PACS number(s): 05.70.Ln, 14.60.Pq, 26.35.+c

\twocolumn

\section{Introduction}
The idea of Gamow, proposed in the 1930s and 1940s~~\cite{Gamow}
about the production of elements through thermonuclear reactions 
in the hot ylem during the early stages of the Universe 
expansion, has been developed during the last 60 years into an elegant 
famous theory of cosmological nucleosynthesis, explaining  
quantitatively the inferred from observational data primordial abundances 
of the light elements~~\cite{CN}.
Thanks to that good accordance between theory
predictions and the observational facts, we nowadays believe to have 
understood well the physical conditions of the nucleosynthesis epoch. 
Still, the uncertainties  of 
the  primordial abundances values extracted from observations 
yet leave a room for physics beyond the standard model. 
 
In this article we present a modification of the standard model 
of cosmological 
nucleosynthesis (CN) - {\it CN with neutrino oscillations}. 
Our aim is twofold: (1) to construct a 
modification of CN using a more precise kinetic approach to the problem of 
nonequilibrium neutrino oscillations and to illustrate the importance of 
such an exact approach, and (2) to determine the cosmologically 
allowed range for oscillation parameters from an accurate study of the 
oscillations effect on the primordial production of helium-4, thus helping 
 clarify the mixing patterns of neutrinos.

The theme of neutrino oscillations is with us almost 
forty years, since the hypothesis for them was proposed by 
Pontecorvo~\cite{pontec}.
 They were studied experimentally and theoretically  
and their cosmological and astrophysical effects have 
been considered in numerous publications~\cite{REV} as far as 
their study  helps to go deeper into the secrets of neutrino 
physics and neutrino mass pattern.
Nowadays there are three main {\it experimental indications} that neutrinos 
 oscillate,
 namely: the solar neutrino deficit~\cite{SUN} (an indirect indication),
the atmospheric neutrino anomaly~\cite{ATM} (an indirect indication)
 and the LSND experiment results~\cite{LSND} (a direct indication). 

(a){\it Solar neutrino deficit:}
  Already four experiments using different techniques 
have detected electron neutrinos from the Sun, at a level significantly lower 
than the predicted on the basis of the Standard Solar Model and the Standard 
Electroweak Theory. Moreover, there exists incompatibility between 
 Chlorine and Kamiokande experiments data, as well as problems 
for predicted berilium and borum neutrinos in the gallium experiments
\cite{SUN,Bah}.
  Recently, it was realized that by changing the  solar model it 
 is hardly possible to solve these problems~\cite{SUNMOD}.
Therefore, it is interesting to find a solution beyond the 
Standard Electroweak
Model. The only  known natural solutions of that kind today are the energy 
dependent MSW neutrino transitions in 
 the Sun interior~\cite{MSW} 
 and the ``just-so'' vacuum oscillations solutions, as well as the recently 
  developed  hybrid solutions of  MSW 
transitions +vacuum oscillations type~\cite{SUNTH}.

(b){\it Atmospheric neutrino anomaly:}
 Three of the five underground experiments on 
atmospheric neutrinos have observed disappearance of muon
neutrinos~\cite{ATM}.
This is in contradiction with the theoretically expected flux of 
muon neutrinos from primary cosmic rays interacting in the atmosphere. A 
successful oscillatory solution of that problem requires large mixing and 
$\delta m^2$ of the order of $10^{-2}$ eV$^2$.

(c){\it Los Alamos LSND experiment} claimed evidence for the oscillation
of $\tilde{\nu}_\mu$
into $\tilde{\nu}_{e}$, with a maximal probability of the order of 
$0.45\times10^{-2}$. A complementary $\nu_{\mu}$ into $\nu_e$
 oscillation search, with completely different systematics and backgrounds, 
 also shows a signal, which indicates the same favoured region of oscillation 
 parameters~\cite{LSND}. 

There exists yet another observational suggestion for  
massive neutrinos and oscillations - the {\it dark matter problem}.
Present models of structure formation in the Universe indicate 
that the observed 
hierarchy of structures is reproduced best by an admixture of 
about 20\% hot dark matter to the cold one~\cite{DM}.
Light neutrinos with  mass  in eV range are the only particle 
dark matter candidates, that are actually known to exist and are 
the most plausible candidates provided by particle physics~\cite{drees}.
 Actually, recent most popular hot plus cold 
dark matter models assume that two nearly degenerate massive neutrinos 
each with mass 2.4 eV play the role of the hot dark matter.
This small mass value is now accessible only by oscillations.

However, in case we take seriously each of these experiments 
pointing to a neutrino anomaly and the neutrino oscillation solution 
to them, a fourth neutrino seems inevitable. 
In the case of only three species of light neutrinos 
with normal interactions and a see-saw hierarchy between the three masses,
it is hardly possible to accommodate all the present data simultaneously.
The successful attempts to reconcile the LSND results with neutrino 
oscillation solutions to the solar and atmospheric neutrino problems 
usually contain some ``unnatural'' features, like forth ultra-light sterile 
neutrino species, or inverted neutrino mass hierarchy~\cite{ALL}.
However, an additional light (with mass less than 1 MeV) 
flavour neutrino is forbidden both from cosmological considerations 
and the experiments on $Z$ decays at LEP~\cite{NLEP}.
Hence, it is reasonable to explore in more detail the 
possibility for an additional light {\it sterile} neutrino.
Besides, GUT theories  ($SO(10), E_6$, etc.) \cite{hr,E6} and SUSY 
theories~\cite{ps,SUSY} predict the existence of a
sterile neutrino. Moreover, recently models of singlet 
fermions, which explain the smallness of sterile neutrino mass 
and its mixing with the usual neutrino were proposed \cite{SUSY}.
 Therefore, it may be very useful to 
obtain more precise information about the cosmologically 
allowed range for the neutrino mixing parameters and thus  present 
 an additional independent test for the already discussed neutrino puzzles. 
 Moreover, the very small values of mass differences, which can  
be explored by the oscillations cosmological effects  
(like the ones discussed 
in our model) are beyond the reach of 
present and near future experiments. 

The present work is a step towards this: we suppose the existence 
of a sterile neutrino ($SU(2)$-singlet)  $\nu_s$, and explore the 
cosmological effect 
of {\it nonresonant neutrino oscillations} $\nu_e \leftrightarrow \nu_s$ 
on the primordial nucleosynthesis,
obtaining thus cosmological constraints on the neutrino mixing parameters.
The nonresonant case in the Early Universe medium corresponds to 
the resonant case in the Sun, therefore, the obtained information is
also of interest for the MSW solution to the solar neutrino problem. 
 We  discuss the special case of {\it nonequilibrium oscillations}
between weak interacting and sterile neutrinos for 
{\it small mass differences} 
 $\delta m^2$, as far as the case of large $\delta m^2$ is already
sufficiently well studied~\cite{bd1}-\cite{s}. 
Oscillations between active and sterile neutrinos,  
effective before neutrino freezing at 2 MeV, 
leading to  $\nu_s$ thermalization before 2 MeV have been studied there. 
T.e. mainly the equilibrium oscillations were considered with rates of 
oscillations and neutrino weak interactions greater 
than the expansion rate.
Here we discuss nonequilibrium oscillations between electron 
neutrinos $\nu_e$ and sterile neutrinos $\nu_s$
for the case when $\nu_s$ do not thermalize till $\nu_e$ decoupling at  
2 MeV and oscillations become effective after $\nu_e$ decoupling.
Such kind of active-sterile neutrino oscillations in vacuum 
 was first precisely studied in~\cite{dpk} using the  
accurate kinetic approach for the description of oscillating 
neutrinos, proposed in the pioneer work of Dolgov~\cite{do}.
However, the thermal background in the prenucleosynthesis epoch 
may strongly affect the propagation of neutrino~\cite{medium,the}
and the account of the neutrino interactions
with the primeval plasma is obligatory~\cite{nr,bd1,ekm}.
The precise kinetic consideration of oscillations in a medium 
was provided in~\cite{our}.
It was proved that in case  when the 
 Universe expansion, the oscillations and the neutrino interactions with the 
medium have comparable rates, their effects should be accounted for 
simultaneously,
using the exact kinetic equations for the neutrino density matrix.
Moreover, for the nonequilibrium oscillations 
energy distortion and asymmetry between neutrinos and antineutrinos may
play a considerable role. As far as both neutrino collisions and active-sterile 
neutrino oscillations distort the initially equilibrium 
active neutrino momentum
distribution, the momentum degree of freedom in the description of neutrino 
must be accounted for. Therefore, for the case of nonequilibrium oscillations 
the evolution of neutrino ensembles should be studied using the exact 
kinetic equations for the {\it density matrix of neutrinos in momentum space}.
 This approach allows an exact investigation of the different effects 
of neutrino oscillations~\cite{dpk,our,sr}: 
depletion of the neutrino number densities, the 
energy distortion and the generation of asymmetry, for each separate momentum
of the neutrino ensembles. 

In the present work we expand the original
investigation~\cite{our}
for the full parameter space of the nonequilibrium oscillations 
model for the nonresonant case. (The resonant case 
will be discussed in a following publication.)
We have provided an exact kinetic
analysis of the neutrino evolution by a numerical integration of 
the {\it kinetic equations
  for the neutrino density matrix for each momentum mode}. 
The kinetic equations are coupled nonlinear and, therefore, an 
analytic solution is hardly possible in the general case 
of oscillations in a medium.  
We have numerically described the evolution 
of the neutrino ensembles from the $\nu_e$ freezing at 
2 MeV till the formation of helium-4.

We have calculated the production of helium-4 in a detail model of primordial
nucleosynthesis, accounting for the direct kinetic effects of oscillations
 on the neutron-to-proton transitions.  
 The oscillations effect on  CN has been
 considered by many
authors~\cite{bd1}-\cite{s},\cite{lan}, ~\cite{sv}-\cite{dhs}. 
However, mainly the excitation 
of an additional degree of freedom due to oscillations (i.e. an increase 
of the effective degrees of freedom $g$)  
and the corresponding increase 
of the Universe expansion rate $H \sim \sqrt g$, leading to an 
overproduction of helium-4
was discussed. The excluded
regions for the neutrino mixing parameters were obtained from the 
requirement (based on the accordance between the theoretically 
predicted and the extracted from observations light elements abundances)
 that the neutrino types should be less than 3.4:
$N_{\nu}<3.4$~\cite{bd1}-\cite{ekm}. A successful account for  
the electron neutrino depletion due to oscillations was first made in
\cite{bd1} and~\cite{ekm}.
In the present work we have precisely calculated the influence of 
oscillations on the primordially produced helium-4 using 
the exact kinetic equations in momentum space for the neutron number density  
and the density matrix of neutrino, instead of their 
particle densities.
 The accurate numerical analysis of oscillations effect on  
helium production within a model of nucleosynthesis
with oscillations, allowed us to account precisely for the 
following important effects of neutrino oscillations: 
neutrino population depletion, distortion of the neutrino spectrum and the
generation of neutrino-antineutrino asymmetry.       
 This enabled  
us to investigate the zone of very small neutrino mass differences up 
to $10^{-11}$ eV$^2$, which has not been reached before. As a 
result, we have obtained constant helium contours in the mass 
difference -- mixing angle plane for the full range of the parameter 
values of our model. No matter what will be the preferred primordial 
helium value, favoured by future observations, it will be possible to 
obtain the excluded region of the mixing parameters using the results of 
this survey. 

The paper is organized as follows. In Section II we 
present the model of nonequilibrium neutrino 
oscillations. In Section III an exact 
analysis of the neutrino evolution using  kinetic equations
  for the neutrino density matrix for each momentum mode is 
provided. The main effects of nonequilibrium oscillations 
are revealed. In Section IV we investigate $\nu_e$ into $\nu_s$ 
oscillations effect on the primordial 
production of helium  using a numerical nucleosynthesis code. 
We discuss the influence of 
nonequilibrium neutrino oscillations, namely 
electron neutrino depletion, neutrino spectrum distortion and 
the generation of neutrino-antineutrino asymmetry on the 
primordial yield of helium-4. The results and conclusions are 
presented in Section V. 

\section{Nonequilibrium neutrino oscillations - the model}

   The model of  nonequilibrium oscillations between weak interacting
electron neutrinos $\nu_e$ and sterile neutrinos $\nu_s$
for the case when $\nu_s$ do not thermalize till $\nu_e$ decoupling at  
2 MeV and oscillations become effective after $\nu_e$ decoupling
is described in detail in~\cite{our}.
The main assumptions are the following:
\begin{itemize}
\item
Singlet neutrinos decouple much earlier, i.e. at a considerably higher 
temperature than the active neutrinos do: 
 $T_{\nu_s}^F > T_{\nu_e}^F$.
\end{itemize}
This is quite a natural assumption, as far as sterile neutrinos 
do not participate into the ordinary weak interactions.
In the models predicting singlet neutrinos, the interactions of 
$\nu_s$ are mediated by gauge bosons with masses 
$M={\cal O}$(1 TeV)~\cite{hr,ps,gm}.
Therefore, in later epochs after their decoupling, their
temperature and number densities are considerably less than
those of the active neutrinos due to the subsequent annihilations
and decays of particles that have additionally heated the 
nondecoupled $\nu_e$ in comparison with the already decoupled $\nu_s$.
\begin{itemize}
\item
We consider oscillations between  $\nu_s$ 
($\nu_s \equiv \tilde{\nu}_L$) and the active
 neutrinos, according to the Majorana\&Dirac ($M\&D$)
mixing scheme~\cite{b} with mixing present just in the electron sector
$\nu_i={\cal U}_{il}~\nu_l$, $l=e,s$:
\footnote{The transitions between different neutrino flavours were 
proved to have negligible effect on the neutrino number densities 
and on primordial nucleosynthesis because of the very slight deviation
from equilibrium in that case $T_f \sim T_f'$ ($f$ is the flavour
index)~\cite{do,lan,dhs}.}
\end{itemize}
$$
\begin{array}{ccc}
\nu_1 & = & c\nu_e+s\nu_s\\
\nu_2 & = & -s\nu_e+c\nu_s,
\end{array}
$$
where $\nu_s$ denotes the sterile electron antineutrino,
$c=\cos(\vartheta)$, 
$s=\sin(\vartheta)$ and $\vartheta$ is the mixing angle in the electron sector,
the mass eigenstates $\nu_1$ and  $\nu_2$ are 
Majorana particles with masses correspondingly $m_1$ and $m_2$.
We consider the nonresonant case $\delta m^2=m_2^2-m_1^2>0$, 
which corresponds in the small mixing angle limit to a sterile neutrino 
heavier than the active one. 

In this model the element of nonequilibrium is introduced by the 
presence of a  small singlet neutrino
density at 2 MeV  $n_{\nu_s} \ll n_{\nu_e}$, when the oscillations 
between $\nu_s$ and $\nu_e$ become effective. In order to provide such
a small singlet neutrino density the sterile neutrinos should  have 
decoupled from the plasma sufficiently early
in comparison to the active ones and should have not regained their 
thermal equilibrium till 2 MeV~\cite{ma,dpk,our}.
 Therefore, as far as the oscillations into $\nu_e$ and the 
following noncoherent scattering off the background 
may lead to the thermalization of  $\nu_s$, 
 two more assumptions are necessary for the nonequilibrium case to 
have place:

\begin{itemize}
\item
Neutrino oscillations should become effective after the decoupling
of the active neutrinos, $\Gamma_{osc}\ge H$ for $T\le 2$ MeV,
which is realizable for $\delta m^2 \le 1.3 \times 10^{-7}$ eV$^2$~\cite{our}. 
\end{itemize}
\begin{itemize}
\item
Sterile neutrinos should not thermalize till 2 MeV when oscillations 
become effective, i.e. the production rate of $\nu_s$ must be smaller than 
the expansion rate. 
\end{itemize}
The problem of sterile neutrino thermalization  
was discussed in the pioneer
work of Manohar~\cite{ma} and in more recent publications
~\cite{bd1}-\cite{ekm}.
 This assumption limits the allowed range of oscillation parameters for
our model:
 $\sin^2(2\vartheta) \delta m^2 \le 10^{-7}$ eV$^2$ \cite{our}.

  We have assumed here that electron neutrinos decouple at 2 MeV.
  However, the neutrino decoupling process is more complicated. It
has been discussed  in literature in detail~\cite{decoupl}. 
Decoupling occurs when the 
neutrino weak interaction rate $\Gamma_w \sim E^2 n_\nu(E)$ becomes 
less than the expansion rate $H \sim \sqrt g T^2$. Really,  for electron
neutrinos this 
happens at about 2 MeV. Nevertheless, due to the fact that weak interaction 
rate is greater at a higher energy,
some thermal contact between neutrinos and high energy 
plasma remains after 2 MeV, especially for the high energy tail of the 
neutrino spectrum.  In case these high energy neutrinos begin to 
oscillate before their decoupling, the account of this dependence of 
decoupling time on the neutrino momentum will be essential for our model. 
Otherwise, in case these neutrinos do not start oscillating 
before decoupling, there will be no harm considering them 
decoupled earlier, as far as they preserve their equilibrium 
distribution anyway due to their extremely small mass. 
In~\cite{our} we have checked that neutrinos from 
high-energy tail start to oscillate much later than they decouple for 
the range of oscillation parameters considered in our model. 
It can easily be understood from the fact that the 
oscillation rate decreases with energy 
 $\Gamma_{osc} \sim \delta m^2/E_\nu$ 
 and, therefore, 
neutrinos with higher energies begin to oscillate later, 
 namely when $\Gamma_{osc}$ exceeds the expansion rate $H \sim \sqrt g T^2$. 
Hence, the precise account for 
the momentum dependence of the decoupling  does not change  the results
of our model but unnecessarily complicates the analysis and leads to 
an enormous increase of the calculation time. Therefore, in what
follows we have assumed a fixed decoupling time instead of 
 considering the real 
decoupling period - i.e. we have accepted that the electron neutrinos 
have completely decoupled at 2 MeV.

 \section{The kinetics of nonequilibrium neutrino oscillations}

The exact kinetic
analysis of the neutrino evolution, discussed in this Section, 
though much more complicated, reveals some 
important features of nonequilibrium oscillations, that cannot be 
caught otherwise. 
 As far as for the nonequilibrium model discussed the rates of expansion
of the Universe, 
neutrino oscillations and neutrino interactions with the medium may 
be comparable, we have used kinetic equations for neutrinos 
accounting {\it simultaneously} for the participation of neutrinos 
into expansion, oscillations and interactions with the medium. All 
possible reactions of neutrinos with the plasma were considered, 
namely: reactions of neutrinos with the electrons, neutrons and protons, 
neutrinos of other flavours, and the corresponding antiparticles, as 
well as self interactions of electron neutrinos.
These equations contain all effects due to first order on $G_F$
medium-induced energy shifts, second order effects due to non-forward 
collisions, and the effects non-linear on the neutrino 
density matrices like neutrino refraction effects in a medium 
of neutrinos. In the case of nonequilibrium oscillations  
the density matrix of neutrinos may considerably differ from its
equilibrium form.~\footnote{When neutrinos are in equilibrium 
their density matrix
has its equilibrium form, namely $\rho_{ij}=\delta_{ij} \exp(\mu/T-E/T)$,
so that one can work with particle densities instead of $\rho$.
In an equilibrium background, the introduction of oscillations
slightly shifts $\rho$ from its diagonal form, due to the extreme
smallness of the neutrino mass in comparison with the characteristic
temperatures and to the fact that equilibrium distribution of massless
particles is not changed by the expansion~\cite{do}.}
Then, for the correct analysis of nonequilibrium 
oscillations, it is important to 
work in terms of 
density matrix of neutrinos in momentum space~\cite{do,dpk,our,sr}.
 Therefore, we have provided a proper kinetic analysis of the 
neutrino evolution using kinetic equations for the 
{\it neutrino density matrix for each 
momentum mode}. 

Hence, the kinetic equations for the density matrix of the nonequilibrium
oscillating neutrinos in the primeval plasma of the Universe
in the epoch previous to nucleosynthesis, i.e. consisting of
photons, neutrinos, electrons, nucleons, 
and the corresponding antiparticles, have the form:
\be
{\partial \rho(t) \over \partial t} =
H p~ {\partial \rho(t) \over \partial p}
+ i \left[ {\cal H}_o, \rho(t) \right]
+i \left[ {\cal H}_{int}, \rho(t) \right]
+ {\rm O}\left({\cal H}^2_{int} \right),
\label{kin}
\ee
where $p$ is the momentum of electron neutrino and $\rho$ is the 
density matrix of the massive Majorana neutrinos in momentum space.

The first term in the equation describes the effect of expansion,
the second is responsible for oscillations, the
third accounts for forward neutrino scattering off the
medium and the last one accounts for second order interaction 
effects of neutrinos with the medium. 
${\cal H}_o$ is the free neutrino Hamiltonian:
$$
{\cal H}_o = \left( \begin{array}{cc}
\sqrt{p^2+m_1^2} & 0 \\ 0 & \sqrt{p^2+m_2^2}
\end{array} \right),
$$
while ${\cal H}_{int} = \alpha~V$ is the interaction Hamiltonian,
where $\alpha_{ij}=U^*_{ie} U_{je}$, 
$V=G_F \left(+L - Q/M_W^2 \right)$,
and in the interaction basis plays the role of an induced squared mass 
for electron neutrinos:

$$
{\cal H}_{int}^{LR} = \left( \begin{array}{cc}
V & 0 \\ 0 & 0 \end{array} \right).
$$

Hence, $V$ is the time varying (due to the Universe cooling) 
effective potential,
induced by the interactions of neutrino with the medium through which it 
propagates. Since $\nu_s$ does not interact with the medium it has no 
self-energy correction, i.e. $V_s=0$.

The first `local' term in $V$ accounts for charged- and neutral-current
tree-level interactions of $\nu_e$ with medium protons, neutrons, 
electrons and positrons, neutrinos and antineutrinos.
It is proportional to
the fermion asymmetry of the plasma $L=\sum_f L_f$, which is 
usually taken to be of the order of the baryon one  i.e. $10^{-10}$
(i.e. $B-L$ conservation is assumed).
$$
L_f \sim {N_f-N_{\bar{f}} \over N_\gamma}~T^3 \sim
{N_B-N_{\bar{B}} \over N_\gamma}~T^3 = \beta T^3.
$$
 The second `nonlocal' term in $V$ arises as an $W/Z$ propagator effect,
$Q \sim E_\nu~T^4$~\cite{nr,bd1}. For the early Universe conditions both 
terms must be accounted for because although the second term is of 
the second power of $G_F$ , the first term is proportional besides 
to the first power of $G_F$, also to the small value of the fermion
asymmetry. Moreover, the two terms have different temperature 
dependence and an interesting interplay between them during the cooling
of the Universe is observed. At high temperature the nonlocal term
dominates, while with cooling of the Universe in the process of 
expansion the local one becomes more important. 

The last term in the Eq.~(\ref{kin}) describes the weak interactions
of neutrinos with the medium. For example, for the weak reactions
of neutrinos with electrons and positrons $e^+ e^- \leftrightarrow
\nu_i \tilde{\nu}_j$, $e^\pm \nu_j \to e'^\pm \nu'_i$ it has the form
$$
\begin{array}{cl}
 & \int {\rm d}\Omega(\tilde{\nu},e^+,e^-)\left[
n_{e^-} n_{e^+} {\cal A} {\cal A}^\dagger - \frac{1}{2} \left\{
\rho,~ {\cal A}^\dagger \bar{\rho} {\cal A} \right\}_+ \right] \\
+ & \int {\rm d}\Omega(e^-,\nu',e'^-)\left[
n'_{e^-} {\cal B} \rho' {\cal B}^\dagger - \frac{1}{2} \left\{
{\cal B}^\dagger {\cal B}, ~\rho \right\}_+ n_{e^-} \right] \\
+ & \int {\rm d}\Omega(e^+,\nu',e'^+)\left[
n'_{e^+} {\cal C} \rho' {\cal C}^\dagger - \frac{1}{2} \left\{
{\cal C}^\dagger {\cal C}, ~\rho \right\}_+ n_{e^+} \right],
\end{array}
$$
where $n$ stands for the number density of the interacting particles,
$$
\begin{array}{cl}
{\rm d}\Omega(i,j,k)&={(2\pi)^4 \over 2E_\nu} 
\int {{\rm d}^3 p_i \over (2\pi)^3~2E_i}{{\rm d}^3 p_j \over (2\pi)^3~2E_j}
{{\rm d}^3 p_k \over (2\pi)^3~2E_k}\\
&\times \delta^4(p_\nu+p_i-p_j-p_k) 
\end{array}
$$
is a phase space factor, 
${\cal A}$ is the amplitude of the process
$e^+ e^- \to \nu_i \tilde{\nu}_j$,
${\cal B}$ is the amplitude of the process
$e^- \nu_j \to e'^- \nu'_i$ and
${\cal C}$ is the amplitude of the process
$e^+ \nu_j \to e'^+ \nu'_i$.
They are expressed through the known amplitudes
${\cal A}_e(e^+ e^- \to \nu_e \tilde{\nu}_e)$,
${\cal B}_e(e^- \nu_e \to e^- \nu_e)$ and
${\cal C}_e(e^+ \nu_e \to e^+ \nu_e)$:
$$
{\cal A} = \alpha~{\cal A}_e,~~~~~
{\cal B} = \alpha~{\cal B}_e,~~~~~
{\cal C} = \alpha~{\cal C}_e.
$$

An analogues equations hold for the antineutrino density matrix,
 the only difference being in  
 the sign of the lepton asymmetry: $L_f$ is replaced by $-L_f$.
 Medium terms depend on neutrino  density, thus introducing a
nonlinear feedback mechanism.
 Neutrino and 
antineutrino ensembles evolve differently as far as the 
background is not CP symmetric. Oscillations may change 
neutrino-antineutrino asymmetry and it in turn affects oscillations. 
The evolution of neutrino and antineutrino ensembles is 
coupled and hence, it must be considered simultaneously. 

We have analysed the evolution of the neutrino density matrix
for the case when  oscillations become noticeable
after electron neutrinos decoupling, i.e. after 2 MeV. 
Then the last term in the kinetic equation can be neglected. 
So, the equation (\ref{kin}) results into a set of coupled nonlinear
integro-differential equations with time dependent coefficients 
for the components
of the density matrix of neutrino. It is convenient 
instead of $\partial / \partial t$  to use  $\partial / \partial \mu$, 
where $\mu^2=\sqrt{16\pi^3 g/45}~(\delta^2/M_{Pl})~t$, and 
$\delta=m_n-m_p$.

Then from eq. (\ref{kin}) we obtain:

\be
\left(\begin{array}{c}
\rho'_{11}\\
\rho'_{22}\\
\rho'_{12}\\
\rho'_{21}\\
\end{array}\right) =
\left(\begin{array}{cccc}
0 & 0 & +iscV & -iscV\\
0 & 0 & -iscV & +iscV\\
+iscV & -iscV & -iM & 0\\
-iscV & +iscV & 0 & +iM\\
\end{array}\right)
\left(\begin{array}{c}
\rho_{11}\\
\rho_{22}\\
\rho_{12}\\
\rho_{21}\\
\end{array}\right),
\label{neutrino}
\ee 

\noindent
 where prime denotes $\partial / \partial \mu$ 
 and  $M=\delta m^2/(2E_\nu)+(s^2-c^2)V$.

Analytical solution is not 
possible without drastic assumptions and, therefore, we have numerically 
explored the problem~\footnote{For the case of vacuum neutrino oscillations
this equation was analytically solved and the evolution of density 
matrix was given explicitly in Ref.~\cite{dpk}.}
 using the Simpson method for integration and the fourth order 
Runge-Kutta algorithm for the solution of the differential equations.

The neutrino kinetics down to 2 MeV does not differ from the 
standard case, i.e. electron neutrinos maintain their equilibrium 
distribution, while sterile neutrinos are absent. So, the initial 
condition for the neutrino ensembles in the interaction basis 
 can be assumed of the form:

$$
{\cal \varrho} = n_{\nu}^{eq}
\left( \begin{array}{cc}
1 & 0 \\
0 & 0 
\end{array} \right)
$$
where $n_{\nu}^{eq}=\exp(-E_{\nu}/T)/(1+\exp(-E_{\nu}/T))$.

 We have analyzed the evolution of nonequilibrium oscillating
neutrinos by numerically integrating the kinetic equations (\ref{neutrino})
 for the period after 
the electron neutrino decoupling till the freeze out of the
neutron-proton ratio
($n/p$-ratio), i.e. for the 
temperature interval $[0.3,2.0]$ MeV.
The oscillation parameters range
studied is $\delta m^2 \in [10^{-11}, 10^{-7}]$ eV$^2$
 and $\vartheta \in[0,\pi/4]$.
  
The distributions of electrons and positrons were taken the
equilibrium ones. 
Really, due to the enormous rates of the electromagnetic reactions 
 of these particles  the deviations from equilibrium are negligible.
We have also neglected the distortion of the neutrino spectra 
due to residual interactions between the electromagnetic and neutrino 
components of the plasma after 2 MeV. This distortion was accurately 
studied in~\cite{dhs}, where it was shown that the relative corrections
to $\nu_e$ density is less than 1 \% and the effect on the primordial 
helium abundance is negligible.
  
The neutron and proton number 
densities, used in the kinetic equations for neutrinos, were substituted  
from the numnerical calculations in CN code accounting for 
neutrino oscillations. I.e. we have simultaneously solved the 
equations governing the evolution of neutrino ensembles and those 
describing the evolution of the nucleons (see the next section). 
The baryon asymmetry $\beta$, parametrized as the ratio of the baryon number 
density to the photon number density,  
was taken to be $3\times 10^{-10}$.

Three {\it main effects of neutrino nonequilibrium oscillations} were
revealed and precisely studied, namely electron neutrino 
depletion, neutrino energy spectrum distortion and the 
generation of asymmetry between neutrinos and their antiparticles:

(a) Depletion of $\nu_e$ population due to oscillations:
 As far as oscillations become effective when the number densities 
of $\nu_e$ are much greater than those of $\nu_s$, 
$N_{\nu_e} \gg N_{\nu_s}$, the oscillations tend to reestablish 
the statistical equilibrium between different oscillating species.
As a result $N_{\nu_e}$ decreases in comparison to its standard
equilibrium value due to oscillations 
in favour of sterile neutrinos.~\footnote{Note, that while neutrinos 
are in thermal equilibrium with the plasma 
 no dilution of their number density  is expected as far as 
it is kept the equilibrium one due to 
the annihilations of the medium electrons and positrons.}
The effect of depletion may be very strong (up to $50\%$) 
 for relatively great $\delta m^2$ and maximal mixing. 
This result of our study is in accordance with other publications 
concerning depletion of electron neutrino population due to oscillations, 
like~\cite{ekt}, however, we have provided more precise account for 
this effect due to the accurate kinetic approach used. 

In Figs.~1 the evolution of neutrino number densities is plotted. 
In Fig.~1a the curves represent the  evolution of the 
electron neutrino number density in the discussed model 
with a fixed mass difference $\delta m^2=10^{-8}$ eV$^2$ and for 
different mixings. The numerical analysis showed that for small mixing,
$\sin^2(2\vartheta)< 0.01$, the 
results do not differ from the standard case, i.e. then oscillations  
may be neglected. In Fig.~1b the evolution of the 
electron neutrino number density is shown  for a nearly maximum mixing,  
$\sin^2(2\vartheta)=0.98$, and different squared mass differences.
Our analysis has proved, that for mass differences 
$\delta m^2<10^{-11}$ eV$^2$, the effect of oscillations is negligible 
for any $\vartheta$.

In case of oscillations effective after the neutrino freeze out, 
electron neutrinos are not in thermal contact with the plasma and, 
therefore, the electron
neutrino state, depleted due to oscillations into steriles, 
 cannot be refilled by electron-positron annihilations. That
 irreversible depletion of $\nu_e$ population exactly equals
 the increase of $\nu_s$ one (see Fig.~1c). The number of the effective 
degrees of freedom do not change due to oscillations in that case, 
as far as the electron 
neutrino together with  the corresponding sterile one contribute to the 
energy density of the Universe as one neutrino unit, even 
in case when the steriles are brought into chemical equilibrium 
with $\nu_e$. This fact was first noted 
in~\cite{ekt}.\footnote{Note the essential difference from the case 
of electron neutrinos in thermal equilibrium, when the oscillations 
into sterile neutrinos bring an 
additional degree of freedom into thermal contact.}

(b) Distortion of the energy distribution of neutrinos:
The effect was first discussed in~\cite{do} for 
 the case of flavour neutrino oscillations. However, as far as the 
energy distortion for that case was shown to be negligible
\cite{do,dhs}, it was 
not paid the necessary attention it deserved. The distortion of the 
neutrino spectrum  was not 
discussed in publications concerning active-sterile neutrino 
oscillations, and was thought to be negligible. In~\cite{dpk} it 
was first shown that for the case of $\nu_e \leftrightarrow \nu_s$ 
 vacuum oscillations this effect is considerable and may even 
exceed that of an additional neutrino species. In~\cite{our} we have 
discussed this effect for the general case of neutrino 
 oscillations in a medium. 
The evolution of the distortion is the following:
Different momentum neutrinos begin to oscillate at different 
temperatures and with different amplitudes.
First the low energy part of the spectrum is distorted, and later on
this distortion concerns neutrinos with higher and higher energies. 
This behaviour is natural, as far as neutrino oscillations affect first 
low energy neutrinos, $\Gamma_{osc} \sim \delta m^2/E_{\nu}$.
The Figs.~2a, 2b, 2c and 2d snapshot the evolution of  
the energy spectrum distortion of active neutrinos $x^2 \rho_{LL}(x)$, 
where $x=E_\nu/T$, for maximal mixing and $\delta m^2 =
10^{-8.5}$ eV$^2$, at different temperatures:
$T=1$ MeV (a), $T=0.7$ MeV (b),  $T=0.5$ MeV (c), $T=0.3$ MeV (d).
As can be seen from the figures, the distortion down to temperatures
of 1 MeV is not significant as far as oscillations are not very
effective and/or the weak residual interactions with the background still
can compensate for the difference.  However, for lower temperatures 
the distortion increases 
and at 0.5 MeV is strongly expressed. Its proper account is important
for the correct determination of oscillations role in the kinetics of 
$n$-$p$ transitions during the freeze out of nucleons at about $0.3$ MeV.

 Our analysis has shown that the account for the nonequilibrium 
distribution by shifting the effective temperature and assuming 
the neutrino spectrum of equilibrium
form, often used in literature (see for example~\cite{ssf}), 
may give misleading results
for the case $\delta m^2 < 10^{-7}$ eV$^2$. The effect cannot be absorbed
merely in shifting the effective temperature and assuming equilibrium 
distributions. For larger neutrino mass differences oscillations are
fast enough and the naive account is more acceptable, provided that 
$\nu_e$ have not decoupled.

(c) The generation of asymmetry between $\nu_e$ and their antiparticles:
The problem of asymmetry generation in different contexts was 
discussed by several authors. The possibility of an asymmetry generation
due to CP-violating flavour oscillations was first proposed in Ref.~\cite{hp}.
Later estimations of an asymmetry due to CP-violating MSW resonant 
oscillations were provided~\cite{la}. The problem of asymmetry was
considered in connection with the exploration of the neutrino
propagation in the early Universe CP-odd plasma also in~\cite{bd1}-\cite{ekm}
 and this type of asymmetry was shown to be negligible.
Recently it was realized in~\cite{ftv,s}, that asymmetry can grow 
to a considerable values for the case of great mass differences, 
$\delta m^2 \ge 10^{-5}$ eV$^2$. The effect of asymmetry for 
small mass differences $\delta m^2 \le 10^{-7}$ eV$^2$ on primordial 
production of helium was also proved to be important for the case of 
resonant neutrino oscillations~\cite{our}.
Our approach allows precise description of the asymmetry evolution, 
as far as working with the {\it self consistent kinetic equations
for neutrinos in momentum space} enables us to calculate the
behaviour of the  asymmetry at each momentum. This is important 
particularly when the distortion of the neutrino spectrum is 
considerable.

In the present work we have explored accurately the effect of the 
asymmetry in the nonresonant case for all mixing angles and for small 
mass differences $\delta m \le 10^{-7}$ eV$^2$. Our analysis showed that 
when the lepton asymmetry is accepted initially equal to the baryon one, 
(as is usually assumed for the popular $L-B$ conserving models), 
the effect of the asymmetry is  small for all the 
discussed parameters range. And although the asymmetry is not wiped out 
by the coupled oscillations, as stated by some authors \cite{ka,ekm}, 
nonresonant neutrino oscillations really cannot generate large
neutrino-antineutrino asymmetry in the early Universe. This result 
is in accordance with the conclusions concerning asymmetry evolution 
in \cite{bd2,ekm,our}. 
We have also checked that the neutrino asymmetry even in the case of 
initial neutrino asymmetry   by two orders of 
magnitude higher does not have significant effect on the
cosmologically produced  $^4\! He$. Therefore, 
 for such small initial values of the lepton 
asymmetry, the neutrino asymmetry should  be better neglected 
when calculating primordial
element production for the sake of computational time.
Mind, however, that for higher values of the initial asymmetry the effect 
could be significant, and should be studied in detail.  The asymmetry 
evolution  and its effect on He-4 production for unusual high initial 
values of the 
lepton asymmetry will be studied elsewhere~\cite{asym}. 

In conclusion, our numerical analysis showed that the nonequilibrium 
oscillations can considerably deplete the number densities of
electron neutrinos (antineutrinos) and distort their energy spectrum.

\section{Nucleosynthesis with nonequilibrium oscillating neutrinos}

As an illustration of the importance of these effects, 
and hence of the proposed approach to the analysis of nonequilibrium 
neutrino oscillations, we discuss their influence on the primordial 
production of $^4\! He$. The effect of oscillations on nucleosynthesis 
has been discussed in numerous publications~\cite{bd1}-\cite{dpk},
~\cite{sv,hp,la}. 
 A detail kinetic
calculation of primordial yield of helium  for the case of the 
nonequilibrium oscillations in vacuum was made in~\cite{dpk}
and the proper consideration accounting for the neutrino forward
scattering processes off the background particles was  done 
in~\cite{our} for some neutrino mixing parameters.~\footnote{
Calculations of helium production within the full Big Bang
Nucleosynthesis code with oscillations were provided also in~\cite{ssf}, 
however, there the momentum degree of freedom of neutrino was not considered
and a simplifying account of the nonequilibrium was 
used - by merely shifting the neutrino effective temperature and 
working in terms of equilibrium particle densities.}
In the present work we calculate precisely the influence of 
oscillations on the 
production of He-4 within a detail numerical CN model with nonresonant 
nonequilibrium neutrino oscillations. 
The analysis of \cite{our} is expanded for the full space of the  mixing
parameters values.

 Working with exact kinetic 
equations for the nucleon number densities and neutrino density matrix 
in momentum space, enables us to analyze the direct influence of 
oscillations onto the kinetics of the neutron-to-proton transfers
and to account precisely for the neutrino depletion, 
neutrino energy distortion and the generation of asymmetry due 
to oscillations. 

Primordial element abundances depend primarily on the neutron-to-proton ratio 
at the weak freeze out ($(n/p)_f$-ratio) of the reactions 
interconverting neutrons and protons : $n+\nu_e \leftrightarrow p+e$ and 
$n+e^+ \leftrightarrow p+\tilde{\nu_e}$. The freeze out occurs when due to the 
decrease of temperature with Universe expansion these weak interaction 
rates  $\Gamma_w \sim E_\nu^2 n_{\nu}$ become comparable and 
less than the expansion rate $H \sim \sqrt{g}~ T^2$. Hence, the 
 $(n/p)_f$-ratio depends on the  effective relativistic degrees of
 freedom $g$ (through the expansion rate) and the neutrino number densities
and neutrino energy distribution (through the weak rates).
  Therefore, we
calculate  accurately the evolution of neutron number density till its 
freeze-out. Further evolution is due to the neutron decays 
$n \rightarrow p+e+\tilde{\nu_e}$ that proceed 
till the effective synthesis of deuterium begins. As far as the 
expansion rate exceeds 
considerably the decay rate for the characteristic period before 
the freeze out, 
decays are not essential. Therefore,  we have  accounted  
for them adiabatically. 

The master equation, describing the evolution of the neutron
 number density in momentum space $n_n$ for the case of oscillating 
neutrinos $\nu_e \leftrightarrow \nu_s$, reads:
\be
\begin{array}{l}
\left(\partial n_n / \partial t \right) 
 = H p_n~ \left(\partial n_n / \partial p_n \right) +\\
 + \int {\rm d}\Omega(e^-,p,\nu) |{\cal A}(e^- p\to\nu n)|^2 \\
~~~~~~~ \times\left[
n_{e^-} n_p (1-\rho_{LL}) - n_n \rho_{LL} (1-n_{e^-})\right] \\
 - \int {\rm d}\Omega(e^+,p,\tilde{\nu}) |{\cal A}(e^+n\to p\tilde{\nu})|^2\\
~~~~~~~\times\left[
n_{e^+} n_n (1-\bar{\rho}_{LL}) - n_p \bar{\rho}_{LL} (1-n_{e^+})\right].
\end{array}
\ee
The first term on the right-hand side describes the effect of expansion 
while the next ones -- the
processes $e^- + p \leftrightarrow n + \nu_e$ and
$p + \tilde{\nu}_e \leftrightarrow e^+ + n$,
directly influencing the nucleon density. It differs from the standard 
scenario one only by the substitution of $\rho_{LL}$ and 
$\bar{\rho}_{LL}$ instead of $n_{\nu}^{eq} = [1-\exp(E_\nu/T)]^{-1}$.
The neutrino and antineutrino density matrices differ
$\bar{\rho}_{LL} \ne \rho_{LL}$, contrary to the standard model, as a 
result of the different reactions with the CP-odd plasma of the 
prenucleosynthesis epoch. We have accounted for the final state Pauli
blocking for neutrinos and electrons.
 
Particle number densities per unit volume are 
expressed as $N = (2\pi)^{-3}\int{\rm d}^3 p~n(p)$.
Performing the integration on the right-hand side of the equation also 
one gets the final equations for the time evolution of the neutron
 number density:

\begin{eqnarray}
&&\left(\partial N_n / \partial t \right)
= -3H N_n + G_F^2~ \frac{g^2_V+3 g^2_A}{\pi^3}~ T^5~ \times 
\nonumber \\
&\times& \Big\{ N_p \int_0^{\infty} [1-\rho_{LL}(x)]~ \frac{{\rm e}^{-x-y}}
{1+{\rm e}^{-x-y}}~
f(x,y){\rm d}x \nonumber \\
&-& N_n\int_0^{\infty} \rho_{LL}(x)~ \frac{1}{1+{\rm e}^{-x-y}}~ f(x,y)
{\rm d}x \nonumber \\
&+& N_p\int_{(1+\zeta)y}^{\infty} \bar{\rho}_{LL}(x)~
\frac{1}{1+{\rm e}^{-x+y}}
f(x,-y){\rm d}x \nonumber \\
&-& N_n\int_{(1+\zeta)y}^{\infty} [1-\bar{\rho}_{LL}(x)]~
\frac{{\rm e}^{-x+y}}{1+{\rm e}^{-x+y}}~ f(x,-y){\rm d}x \Big\}
\end{eqnarray}
where $f(x,y)=x^2(x+y)\sqrt{(x+y)^2+\zeta^2y^2}$ and
$y=(\delta + m_e)/T$, $\zeta=m_e/\delta$, $\delta = m_n - m_p$.  

The first term on the right-hand side describes the dilution effect of 
expansion, the next describe the weak processes, as pointed above. 
 We have numerically integrated this equation for the temperature range 
of interest $T \in [0.3,2.0]$ MeV for the full range of oscillation 
parameters of our model. The value of $\rho_{LL}(x)$ at each integration step 
was taken from the simultaneously performed integration of the 
set of equations (\ref{neutrino}), i.e. the evolution of neutrino and the
nucleons was followed self consistently. 
The initial values at $T=2$ MeV for the neutron, proton and electron 
number densities are their equilibrium values. 
Although the electron mass is comparable with the temperature in the 
discussed temperature range, the deviation of the electron density from
its equilibrium value is negligible due to the enormous rate of the 
reactions with the plasma photons~\cite{do}.
The parameters values of the CN model, adopted
in our calculations,  are the following: the mean neutron lifetime is 
$\tau= 887$ sec, which corresponds to the present weighted average 
value~\cite{partdata}, the effective number of relativistic flavour types of 
 neutrinos 
during the nucleosynthesis epoch $N_{\nu}$ is assumed equal to the 
standard value $3$.
This is a natural choice as far as it is in good agreement both with 
the CN arguments~\cite{N}~\footnote{However mind also the possibilities 
for somewhat relaxation of 
that kind of bound in modifications of the CN model with decaying particles 
as in~\cite{Nvar,ki}.} and with the precision measurements of the $Z$ decay 
width at LEP~\cite{NLEP}.

\section{Results and conclusions}

The results of the 
numerical integration are illustrated in Fig.~3. 
As it can be seen from the figure the kinetic effects
 (neutrino population depletion and distortion of neutrino spectrum)
 due to oscillations play an important
role and lead to a considerable overproduction of helium.

Qualitatively the effect of oscillations on helium production 
can be described as follows:

The depletion of the electron neutrino number densities due to
oscillations into sterile ones strongly affects the 
$n \leftrightarrow p$ reactions rates. It leads to an effective
decrease in the
 processes rates, and hence to an increase of the freezing
temperature of the $n/p$-ratio and the corresponding overproduction of the
primordially produced $^4\! He$. 

The effect of the distortion of the energy distribution of neutrinos 
has two aspects. On one hand an average decrease of the energy of 
active neutrinos leads to a decrease of the weak reactions rate, 
$\Gamma_w \sim E_\nu^2$ and subsequently to an increase in the 
freezing temperature and the produced helium. On the other hand, 
there exists an energy threshold for the reaction 
$\tilde{\nu}_e+p \to n+e^+$. And in case when, due to oscillations, 
the energy of the relatively greater part of neutrinos becomes 
smaller than that threshold the $n/p$- freezing ratio decreases 
leading to a corresponding decrease of the primordially produced 
helium-4~\cite{ki}. The numerical analysis showed that the 
latter effect is less noticeable compared with the former ones.

The asymmetry calculations showed a slight predominance of neutrinos
over antineutrinos, not leading to a noticeable effect on the 
production of helium in case the lepton asymmetry is accepted initially 
equal to the baryon one. So, the effect of asymmetry is proved to be 
negligible for all the discussed parameter range, i.e. for any 
$\vartheta$ and for $\delta m^2 \le 10^{-7}$ eV$^2$. 
We have partially (not for the full range of model parameters) 
 investigated the problem  for higher than the baryon one 
initial lepton asymmetry. The preliminary results point that even 
lepton asymmetry initially by two orders of magnitude higher does 
not have noticeable  effect on the cosmologically produced  $^4\! He$. 
Higher than those lepton asymmetries, however, should be accounted for
properly even in the nonresonant case. 

Thus, the total result of nonequilibrium neutrino oscillations is
an overproduction of helium in comparison to the standard value.

In Fig.~4  the dependence of the frozen neutron
number density relative to nucleons $X_n=N_n/(N_p+N_n)$ on the
mixing angle for  different  fixed  $\delta m^2$ is illustrated. 
 The dependence of the frozen neutron
number density relative to nucleons $X_n=N_n/(N_p+N_n)$ on the
$\delta m^2$ for fixed different mixing angles, is presented in Fig.~5.
The effect of oscillations is maximal at maximal mixing for the 
nonresonant case of neutrino oscillations. 
 As it can be seen from the figures, it 
becomes almost negligible (less than $1\%$) for mixings as small as 0.1
for any $\delta m^2$ of the discussed range of our model.
The value of the frozen $n/p$-ratio is a smoothly increasing function of the 
mass difference. 
Our analysis shows that the effect of oscillation 
for $\delta m^2$  smaller than  $10^{-10}$ eV$^2$ even for maximal 
mixing is  smaller than $1\%$. The  
nonresonant oscillations with $\delta m^2 \le 10^{-11}$ eV$^2$ do not 
have any observable effect on the primordial production of elements, i.e. the 
results coincide with the standard model values with great accuracy.

From the numerical integration for different oscillation parameters
we have obtained the primordial helium yield 
$Y_p(\delta m^2,\vartheta)$, which is illustrated by the 
surface in Fig.~6. Some of the  constant helium 
contours calculated in the discussed model of cosmological 
nucleosynthesis with nonresonant neutrino oscillations 
on the $\delta m^2-\vartheta$ plane are presented in Fig.~7. 

 On the basis of these results, requiring an agreement between the 
 theoretically predicted and the observational values of helium, 
  it is possible to obtain cosmological 
constraints on the neutrino mixing parameters. 
At present the primordial helium values 
extracted from  observations  differ considerably: 
for example some authors believe that the systematic errors have already 
been reduced to about the same level as the statistical one and 
obtain the bounds for the primordial helium:
$Y_p(^4\! He)=0.232\pm 0.003$~\cite{Ystand},
 while others argue that underestimation of the systematic errors, such
as errors in helium emissivities, inadequatisies in the radiative 
transfer model used,  corrections for underlying stellar absorption 
and fluorescent enhancement in the He~I lines, corrections for neutral 
helium, may be significant and their account may raise the upper bound 
on $Y_p$ as high as 0.26~\cite{Ynstand}. Thus besides the widely adopted 
 ``classical'' bound $Y_p<0.24$~\cite{Yclas} it is reasonable to have in 
mind the more ``reliable'' upper bound to the primordial helium abundance
 $Y_p<0.25$ and even the extreme value as high as 0.26~\cite{Ynclas}.
Therefore, we considered it useful to provide the precise calculations 
for helium contours up to  0.26. 
So, whatever the primordial abundance of $^4\! He$ will be found to be 
in future (within this extreme range) the results of  our
 calculations may  provide the corresponding bound on mixing parameters
of neutrino for the case of nonresonant active-sterile oscillations 
with small mass differences. 
Assuming the conventional observational bound on primordial $^4\! He$
$0.24$ 
 the cosmologically excluded  region for the oscillation parameters  
is shown on the plane $\sin^2(2\vartheta)$ - $\delta m^2$ in Fig.~7.
 It is situated to the right of  the $Y_p=0.245$ curve, which 
gives $5\%$  overproduction of helium in comparison with the accepted 
0.24 observational value. 

The curves, corresponding  
to helium abundance $Y_p=0.24$, obtained in the present work, and   
in previous works, analyzing the nonresonant active-sterile neutrino 
oscillations, are plotted in Fig.~8. 
In~\cite{bd1} and~\cite{ekt} the authors estimated the effect of  
excitement of an additional degree of freedom due to oscillations, 
and the corresponding increase 
of the Universe expansion rate, leading to an overproduction of helium-4.
The excluded
regions for the neutrino mixing parameters were obtained from the 
requirement that the neutrino types should be less than 3.4:
$N_{\nu}<3.4$. In these works the depletion effect was considered.
The asymmetry was neglected and the distortion of the neutrino spectrum
was not studied as far as the kinetic equations for neutrino mean 
number densities were considered.  
Our results are in good accordance with the estimations in  
~\cite{bd1} and the numerical analysis in~\cite{ekt}, who have made 
very successful account for one of the discussed effects of 
nonequilibrium oscillations - the neutrino population depletion. 
The results of~\cite{ssf}, as can be seen  from the Fig.~8, 
differ more both from the ones of the previously cited works and from 
our results. Probably 
the account for nonequilibrium oscillations merely by
shifting the effective neutrino temperature, as assumed there is 
not acceptable for a large range of model parameters. 

As can be seen from the curves, for large mixing angles, 
we exclude  $\delta m^2\ge 10^{-9}$ eV$^2$, which is almost an order 
of magnitude stronger constraint than the previously existing. 
This more stringent constraints obtained in our work for the region of 
great mixing angles and small mass differences is due to 
the more accurate kinetic approach we have used  
 and to the precise 
account of neutrino depletion, energy distortion  and asymmetry
due to oscillations. 

As far as we already have at our disposal some impressive indications for 
neutrino oscillations, it is interesting to compare our results 
also with the range of parameters which could eventually 
explain the observed neutrino anomalies:

The vacuum oscillation interpretation of the solar neutrino problem requires 
extremely small mass differences squared,
less and of the order of 
$10^{-10}$ eV$^2$.  It is safely lower than the excluded region, 
obtained in our work, and is, therefore, allowed from CN considerations. 
The MWS small mixing angle nonadiabatic solution (see for example 
Krastev, Liu and Petcov in~\cite{SUNTH})
is out of the reach of our model. However, as we are in a good accordance with 
the results of active-sterile neutrino oscillation models with higher 
mass differences, it is obvious that a natural extrapolation of our 
excluded zone towards higher mass differences will rule out  partially 
the possible solution range for large mixing angles.

Our pattern of neutrino mixing is compatible with models of degenerate 
neutrino masses of the order of 2.4 eV, necessary for the successful 
modelling of the structure formation of the Universe in Hot plus 
Cold Dark Matter Models~\cite{DM}. 
 
 As a conclusion, we would like to outline the main achievements of this work:
In a model of nonequilibrium nonresonant active-sterile oscillation, we had 
studied the effect of oscillations on the evolution of the neutrino number 
densities, neutrino spectrum distortion and neutrino-antineutrino asymmetry.
 We  have used kinetic equations for the density matrix of neutrinos in
{\it momentum} space, accounting {\it simultaneously} for expansion,
 oscillations and interactions with the medium. This approach enabled us to 
 describe precisely the  behaviour of neutrino ensembles in
the Early Universe in the period of interest for CN. 
The analysis was provided for small mass 
differences. We have
shown that the energy distortion may be significant, while the asymmetry
 in case it is initially (i.e. before oscillations become effective) of the 
order of the baryon one, may be neglected. 
 
Next, we have made a precise survey of the influence of the discussed type of 
oscillations on the cosmological production of helium-4. We have 
calculated the
evolution of the corresponding neutron-to-proton ratio from the time 
of freeze out of neutrinos at 2 MeV till the effective freeze out of
nucleons at 0.3 MeV for the full range of model parameters. 
As a result we have obtained the dependence $Y_p(\delta m^2, \vartheta)$ 
and constant helium
contours  on the $\delta m^2 - \vartheta$ plane. Requiring an agreement
between the observational and the theoretically predicted primordial 
helium abundances, we
have calculated accurately  the excluded regions for the neutrino mixing
parameters,
for different assumptions about the preferred primordial value of helium.

\section*{Acknowledgements}
The authors thank prof. A. Dolgov for  useful discussions and encouragement.
 D.K. is grateful to prof. I. Novikov  and prof. P. Christensen for 
the opportunity to work at the Theoretical Astrophysics Center.
 She is glad to thank the Theoretical Astrophysical Center for the warm 
 hospitality and financial support. 
She acknowledges  the hospitality and support of the Niels Bohr 
Institute. This work was supported also by 1996/1997 Danish Governmental 
Scholarship grant.
M.C. thanks NORDITA for the hospitality.  

This work was supported in part by the Danish National Research
Foundation through its establishment of the Theoretical Astrophysics Center.

\vspace{0.2cm}

\pagebreak[1]


\newpage
\onecolumn

\begin{center}{\Large Figure Captions}
\end{center}
\ \\

{\bf Figure 1a}: The curves represent the calculated evolution of the 
electron neutrino number density in the discussed model of 
active-sterile neutrino oscillations with a mass difference 
$\delta m^2=10^{-8}$ eV$^2$ and different mixing, parametrized by 
$\sin^2(2\vartheta)$, namely: $1$, $10^{-0.01}$, $10^{-0.1}$ and 0.1.

\ \\

{\bf Figure 1b}: The curves show the evolution of the 
electron neutrino number density in the discussed model of 
nonresonant active-sterile neutrino oscillations for a nearly maximum mixing,  
$\sin^2(2\vartheta)=0.98$, and different squared mass differences
$\delta m^2$, namely $10^{-7}$,  $10^{-8}$, $10^{-9}$ and $10^{-10}$  
in eV$^2$.

\ \\

{\bf Figure 1c}: The curves show the evolution of the 
electron neutrino number density (the solid curve) and the sterile 
neutrino number density (the dashed curve) in the case of  the 
nonresonant active-sterile neutrino oscillations for a maximal mixing
and  $\delta m^2=10^{-8}$ eV$^2$. The reduction of the active 
neutrino population is exactly counterbalanced by a corresponding 
increase in the sterile neutrino population.

\ \\

{\bf Figure 2}: The figures illustrate the evolution of  the energy spectrum 
distortion of active neutrinos $x^2 \rho_{LL}(x)$, where $x=E_\nu/T$, 
for the case of nonresonant $\nu_e$-$\nu_s$ oscillations with a maximal mixing 
and $\delta m^2 = 10^{-8.5}$ eV$^2$, at different temperatures:
$T=1$ MeV (a), $T=0.7$ MeV (b),  $T=0.5$ MeV (c), $T=0.3$ MeV (d).

\ \\

{\bf Figure 3}: The evolution of the neutron
number density relative to nucleons $X_n(t)=N_n(t)/(N_p+N_n)$ 
 for the case of nonresonant oscillations  with maximal 
mixing and different $\delta m^2$ is shown. For comparison the standard 
model curve is plotted also.

\ \\

{\bf Figure 4}: The figure illustrates the dependence of the frozen neutron
number density relative to nucleons $X_n=N_n/(N_p+N_n)$ on the
mixing angle for  different $\delta m^2$.

\ \\

{\bf Figure 5}: The figure illustrates the dependence of the frozen neutron
number density relative to nucleons $X_n=N_n/(N_p+N_n)$ on the
mass difference  for  different  mixing angles.

\ \\

{\bf Figure 6}: The dependence of the primordially produced helium on 
the oscillation parameters is represented by the surface 
$Y_p(\delta m^2,\vartheta)$.

\ \\

{\bf Figure 7}: On the $\delta m^2-\vartheta$ plane some of the
constant helium contours calculated in the discussed model of 
cosmological nucleosynthesis with nonresonant neutrino oscillations 
are shown.

\ \\

{\bf Figure 8}: The curves, corresponding  
to helium abundance $Y_p=0.24$, obtained in the present work and 
in previous works, analyzing the 
nonresonant active-sterile neutrino oscillations, are plotted on 
the $\delta m^2-\vartheta$ plane. 

\pagestyle{empty}
\newpage

\epsfbox[100 170 700 770]{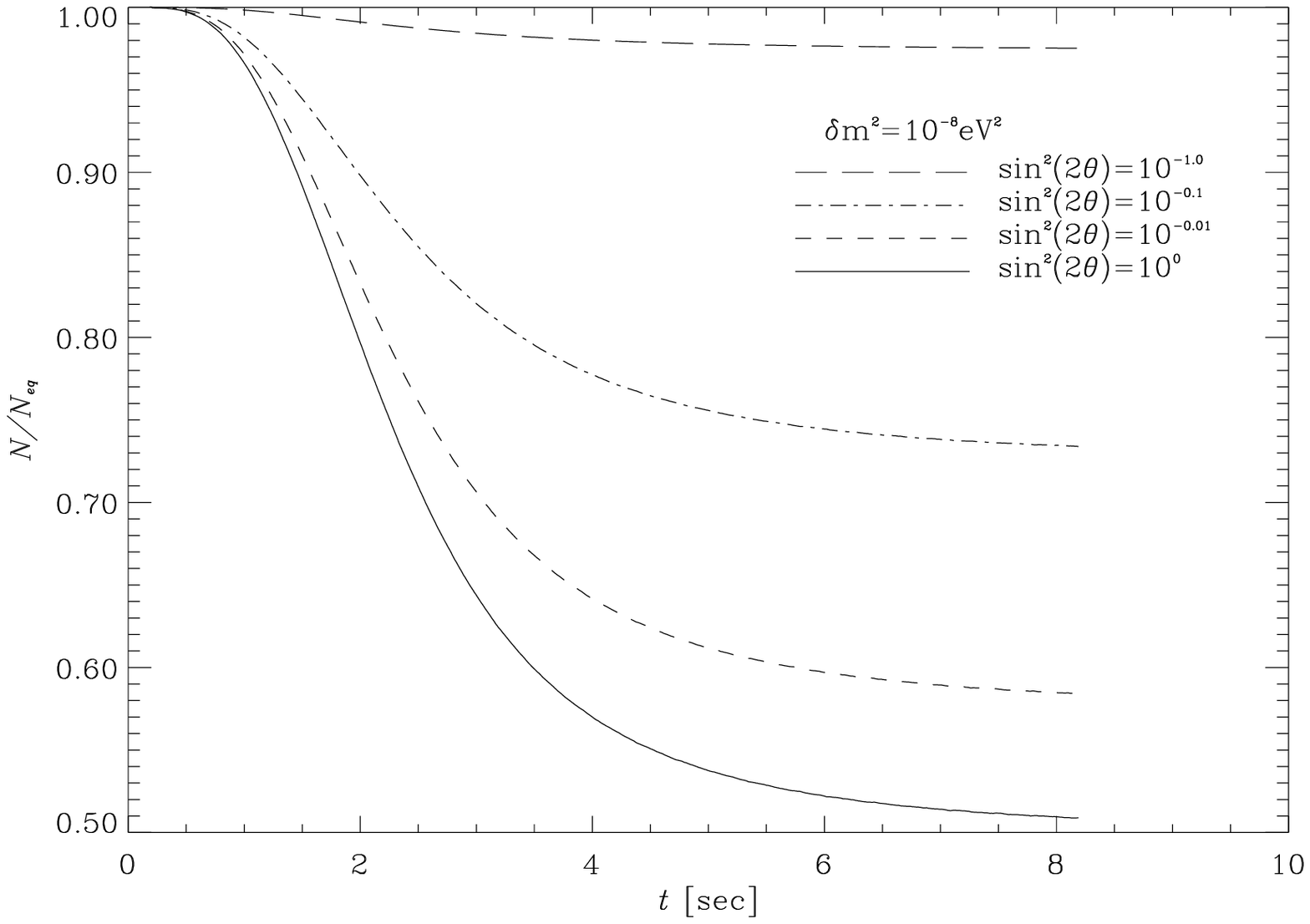}

\vfill

{\bf Figure 1a}: The curves represent the calculated evolution of the 
electron neutrino number density in the discussed model of 
active-sterile neutrino oscillations with a mass difference 
$\delta m^2=10^{-8}$ eV$^2$ and different mixing, parametrized by 
$\sin^2(2\vartheta)$, namely: $1$, $10^{-0.01}$, $10^{-0.1}$ and 0.1.

\newpage

\epsfbox[100 170 700 770]{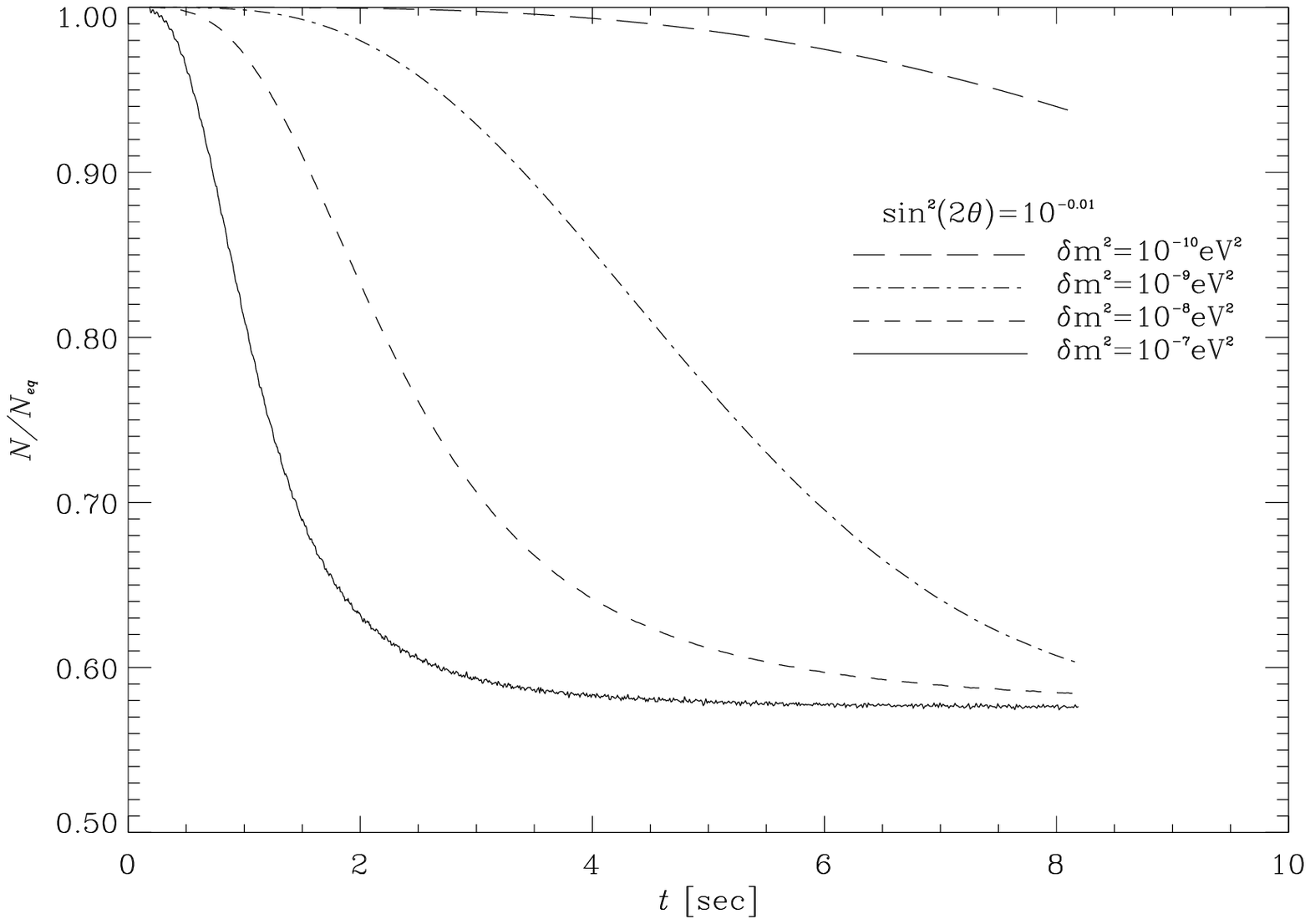}

\vfill

{\bf Figure 1b}: The curves show the evolution of the 
electron neutrino number density in the discussed model of 
nonresonant active-sterile neutrino oscillations for a nearly maximum mixing,  
$\sin^2(2\vartheta)=0.98$, and different squared mass differences
$\delta m^2$, namely $10^{-7}$,  $10^{-8}$, $10^{-9}$ and $10^{-10}$  
in eV$^2$.

\newpage

\epsfbox[100 170 700 770]{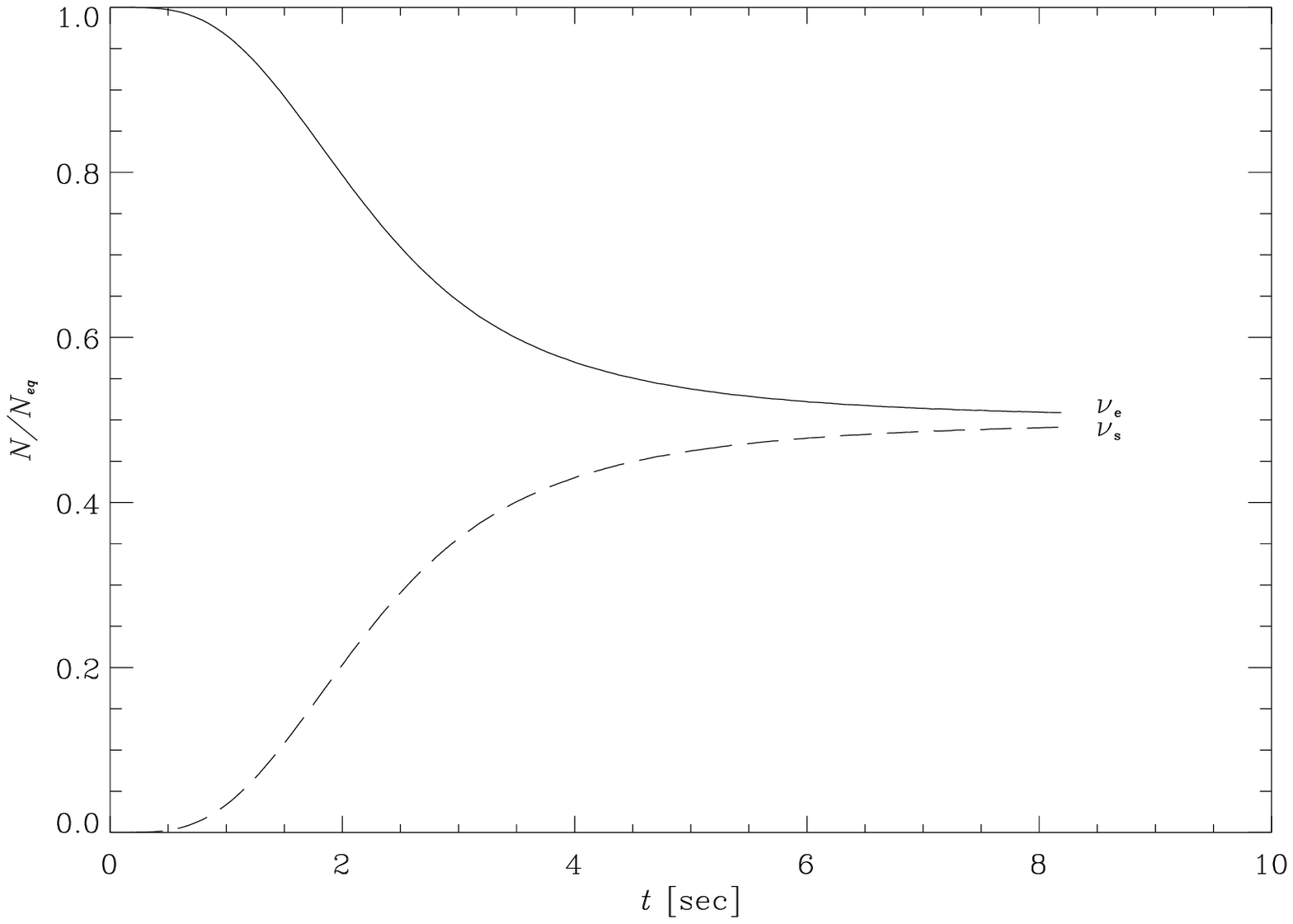}

\vfill

{\bf Figure 1c}: The curves show the evolution of the 
electron neutrino number density (the solid curve) and the sterile 
neutrino number density (the dashed curve) in the case of  the 
nonresonant active-sterile neutrino oscillations for a maximal mixing
and  $\delta m^2=10^{-8}$ eV$^2$. The reduction of the 
active neutrino population is exactly counterbalanced by a 
corresponding increase in the sterile neutrino population.

\newpage

\epsfbox[100 170 700 770]{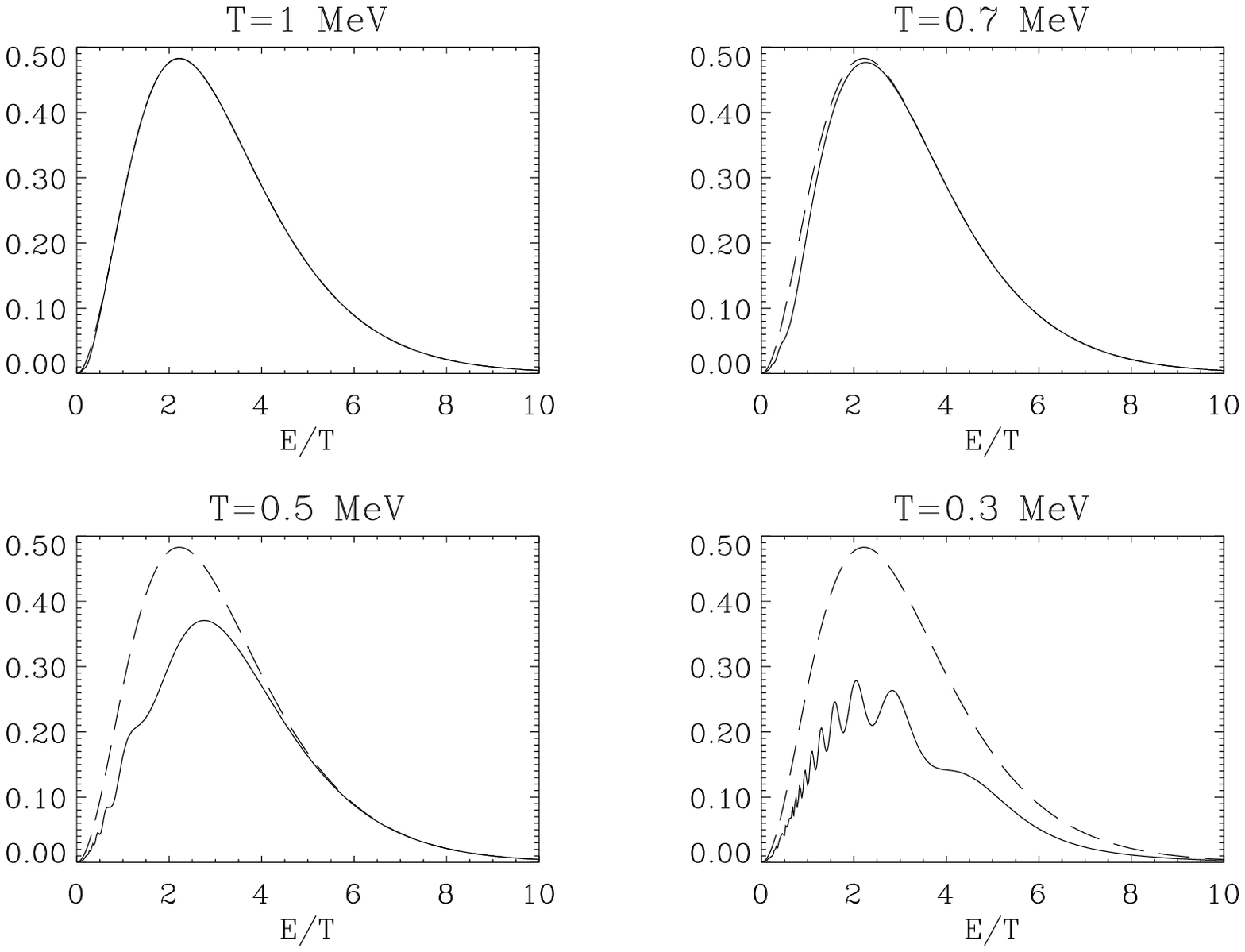}

\vfill

{\bf Figure 2}: The figures illustrate the evolution of  the energy spectrum 
distortion of active neutrinos $x^2 \rho_{LL}(x)$, where $x=E_\nu/T$, 
for the case of nonresonant $\nu_e$-$\nu_s$ oscillations with a maximal mixing 
and $\delta m^2 = 10^{-8.5}$ eV$^2$, at different temperatures:
$T=1$ MeV (a), $T=0.7$ MeV (b),  $T=0.5$ MeV (c), $T=0.3$ MeV (d).

\newpage

\epsfbox[100 170 700 770]{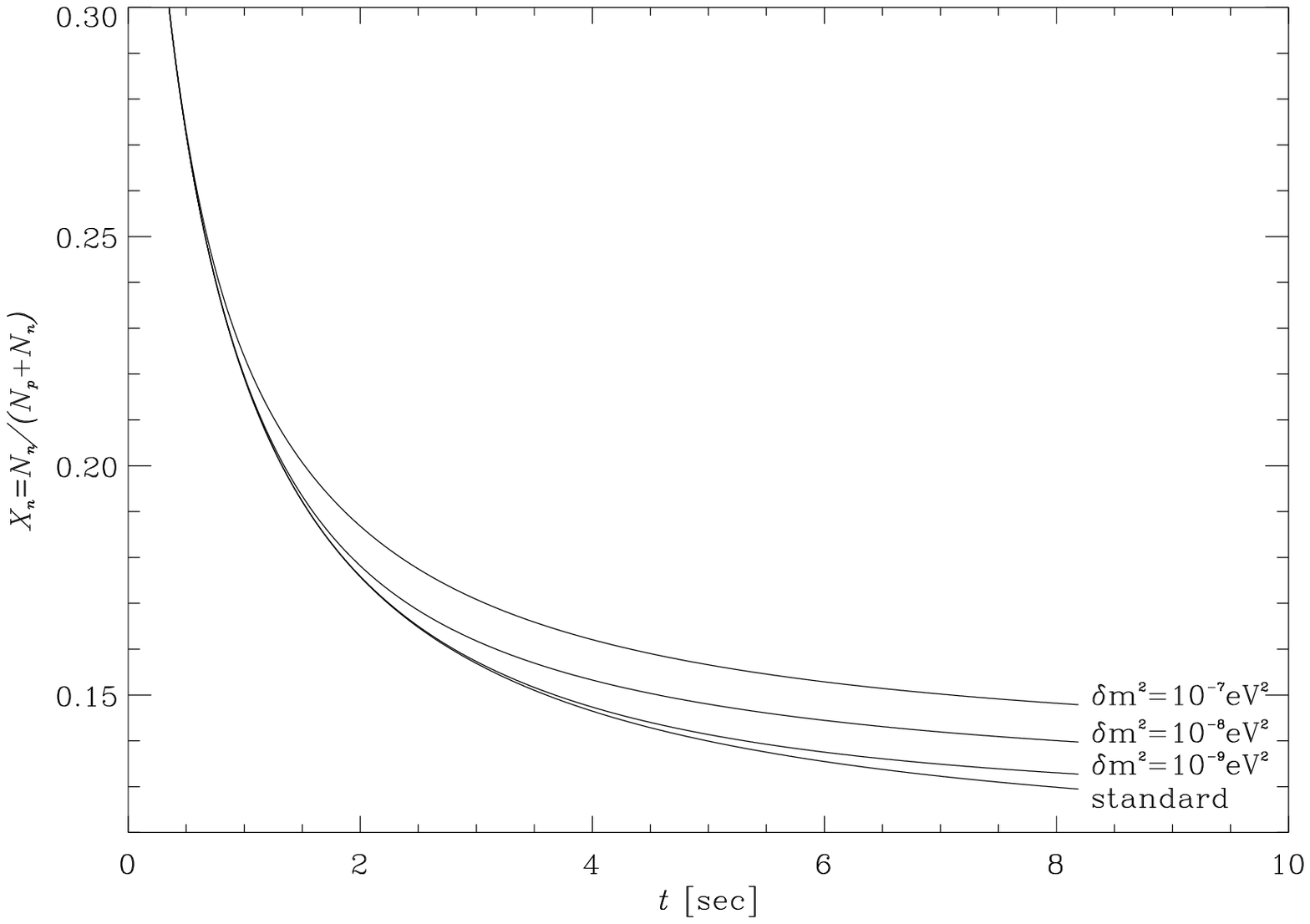}
 
\vfill

{\bf Figure 3}: The evolution of the neutron
number density relative to nucleons $X_n(t)=N_n(t)/(N_p+N_n)$ 
 for the case of nonresonant oscillations  with maximal 
mixing and different $\delta m^2$ is shown. For comparison the standard 
model curve is plotted also.

\newpage

\epsfbox[100 170 700 770]{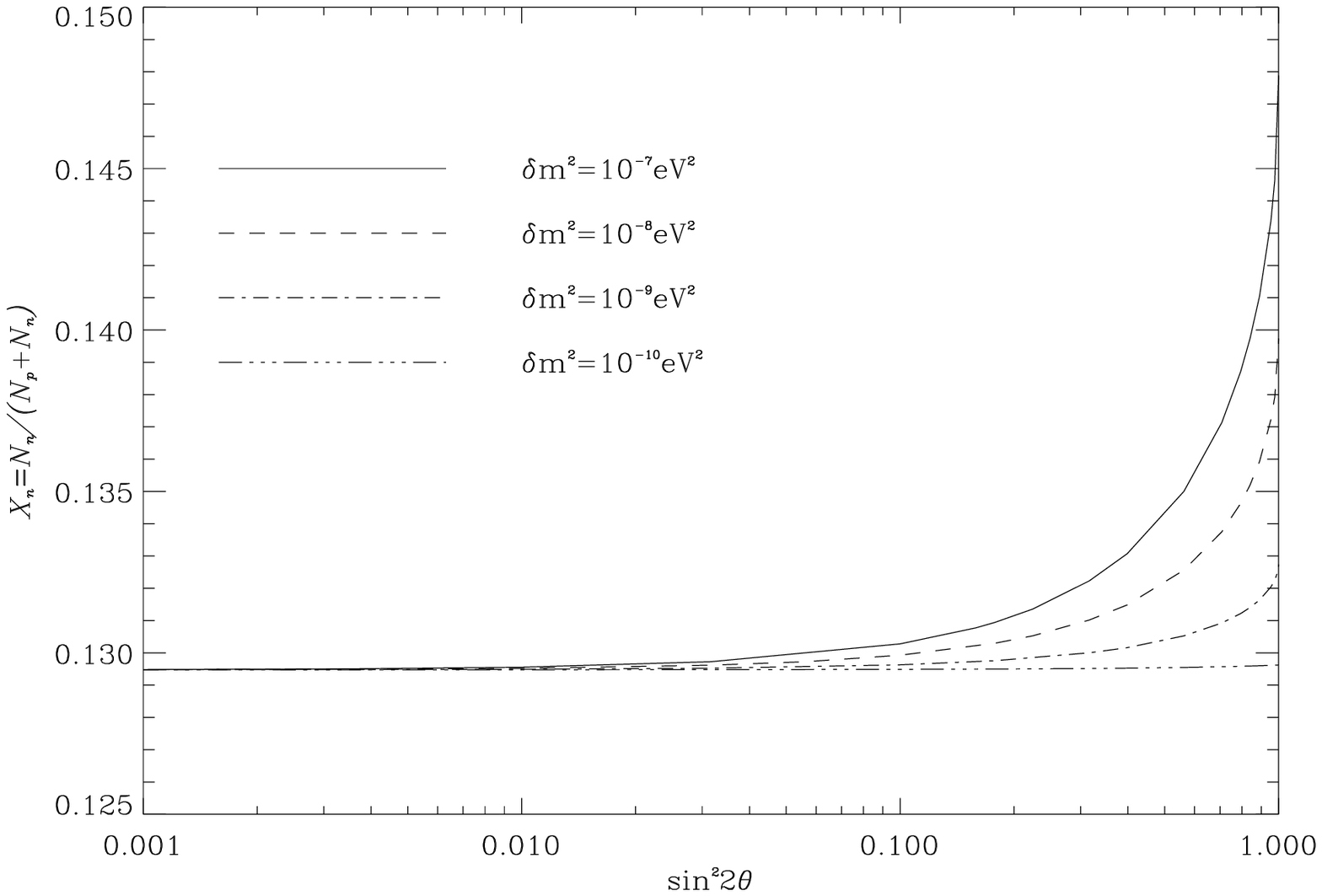}

\vfill

{\bf Figure 4}: The figure illustrates the dependence of the frozen neutron
number density relative to nucleons $X_n=N_n/(N_p+N_n)$ on the
mixing angle for  different $\delta m^2$.

\newpage

\epsfbox[100 170 700 770]{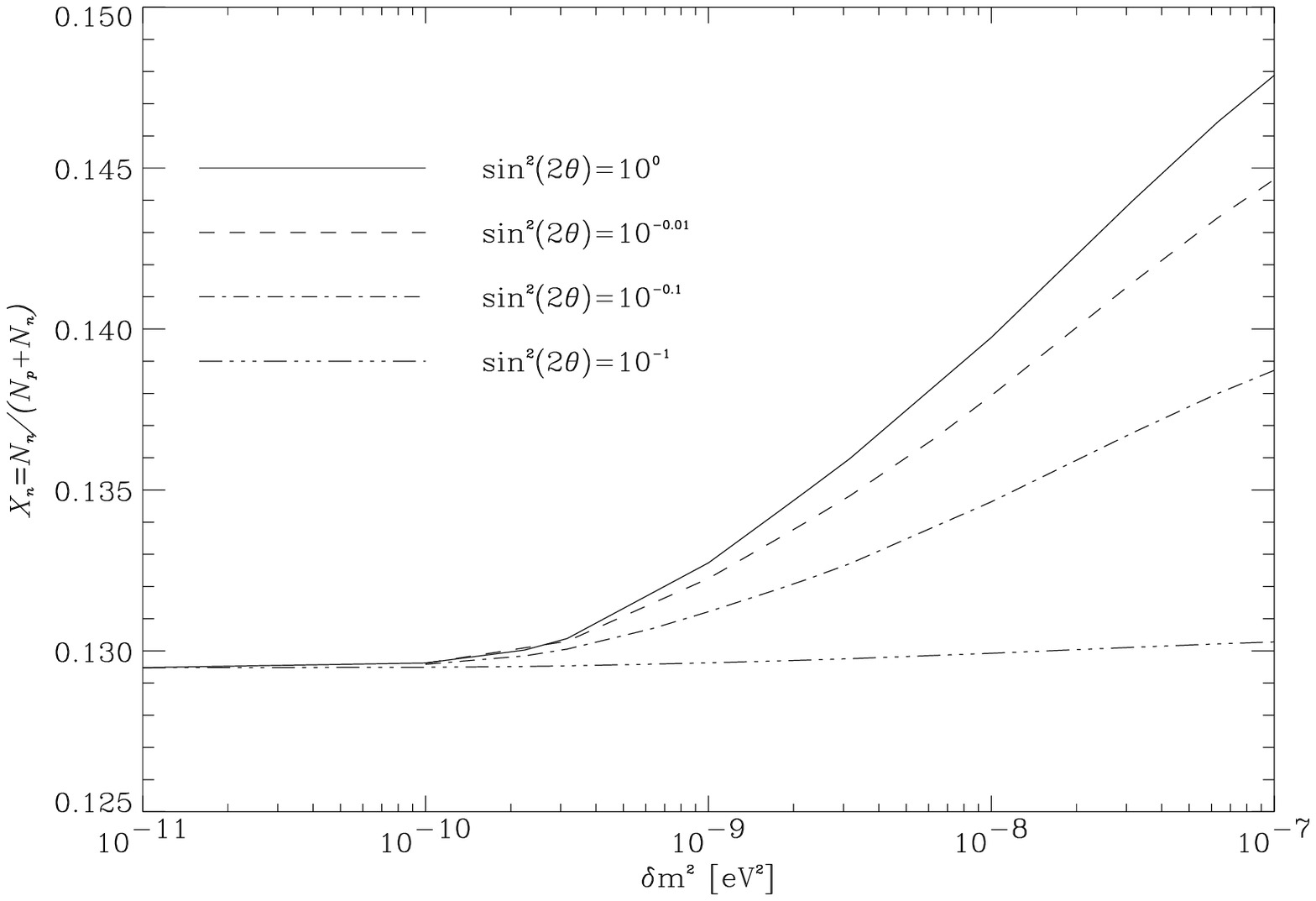}

\vfill

{\bf Figure 5}: The figure illustrates the dependence of the frozen neutron
number density relative to nucleons $X_n=N_n/(N_p+N_n)$ on the
mass difference  for  different  mixing angles.

\newpage

\epsfbox[100 170 700 770]{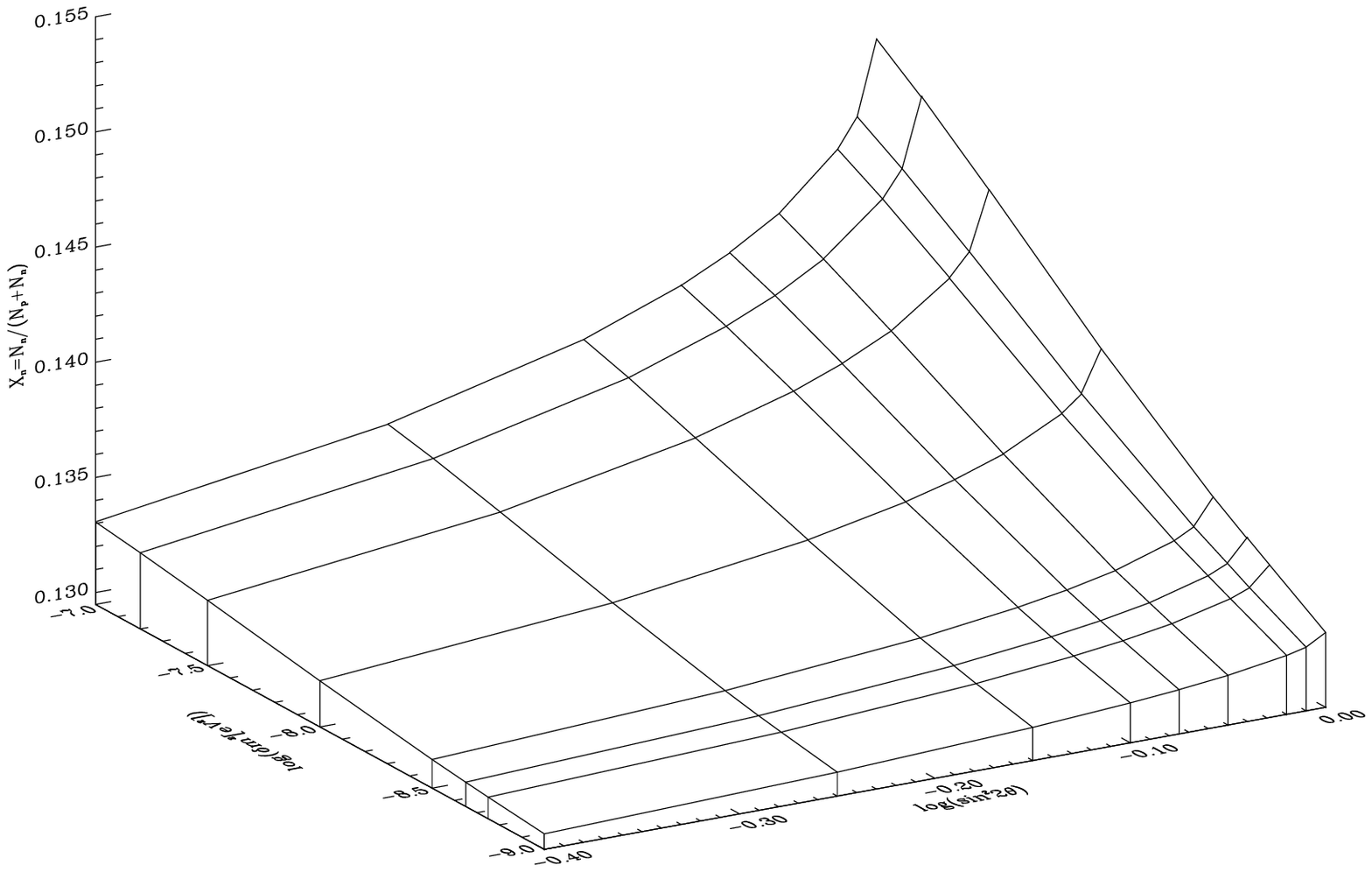}

\vfill

{\bf Figure 6}: The dependence of the primordially produced helium on 
the oscillation parameters is represented by the surface 
$Y_p(\delta m^2,\vartheta)$.

\newpage

\epsfbox[100 170 700 770]{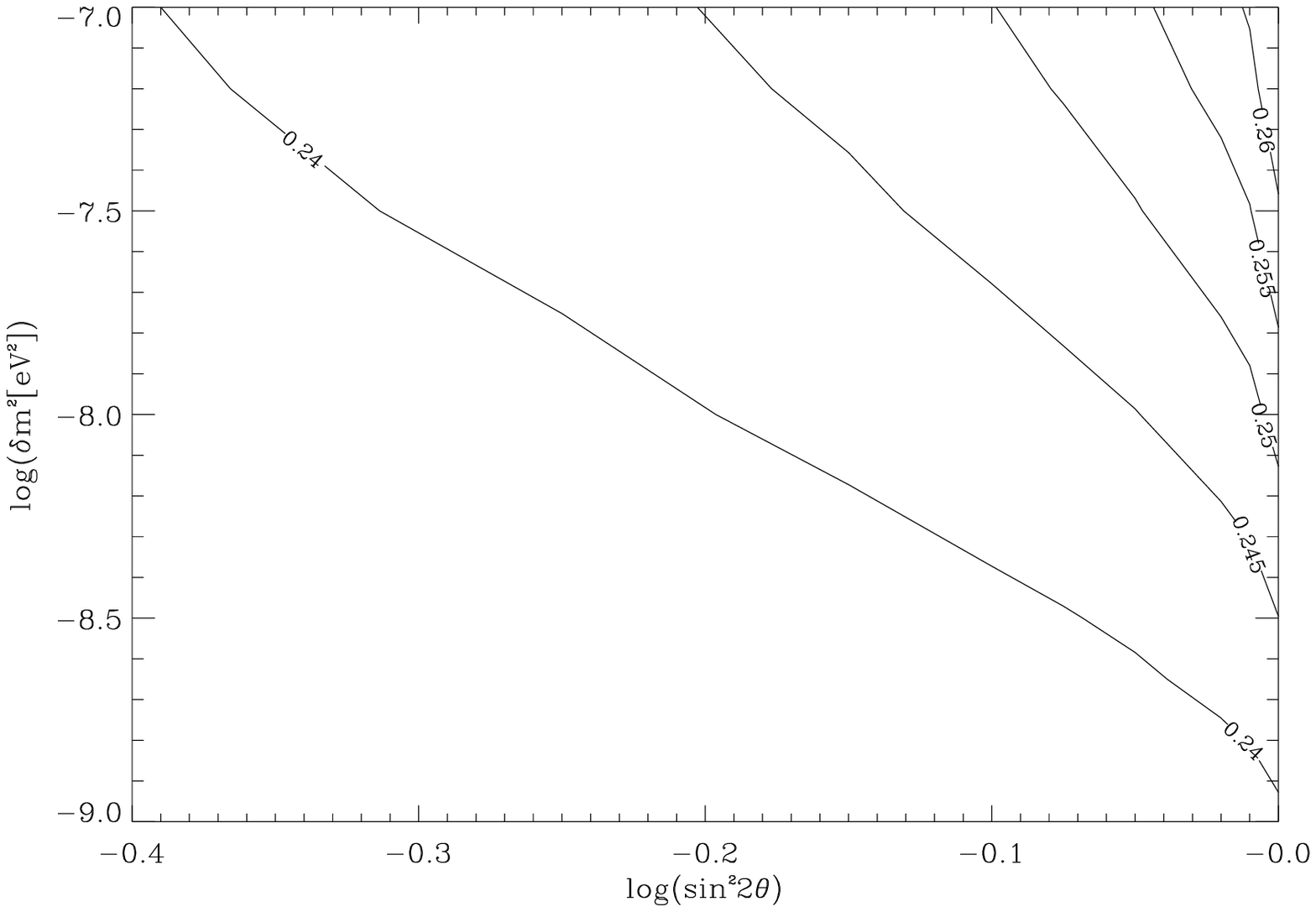}

\vfill

{\bf Figure 7}: On the $\delta m^2-\vartheta$ plane some of the
constant helium contours calculated in the discussed model of 
cosmological nucleosynthesis with nonresonant neutrino oscillations 
are shown.

\newpage

\epsfbox[100 170 700 770]{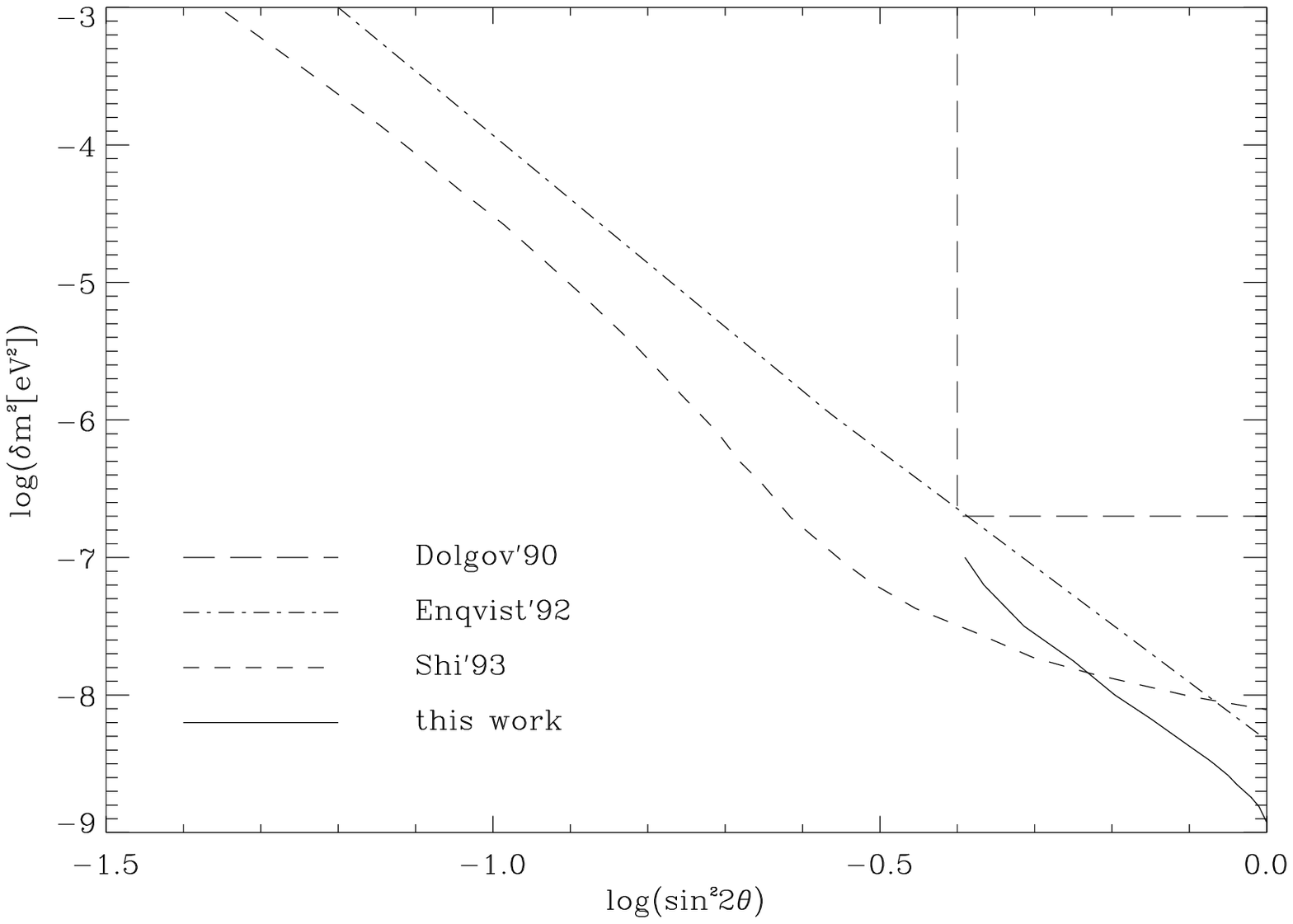}

\vfill

{\bf Figure 8}: The curves, corresponding  
to helium abundance $Y_p=0.24$, obtained in the present work and 
in previous works, analyzing the 
nonresonant active-sterile neutrino oscillations, are plotted on 
the $\delta m^2-\vartheta$ plane. 

\end{document}